\documentclass{article}
\newcommand{\be}{\begin{equation}}
\newcommand{\ee}{\end{equation}}
\def\bea{\begin{eqnarray}}
\def\eea{\end{eqnarray}}
\def\bean{\begin{eqnarray*}}
\def\eean{\end{eqnarray*}}

\newcommand{\p}{\partial}
\newcommand{\dab}{\boldsymbol\delta}
\newcommand{\qbar}{\ov{q}{}}
\newcommand{\cQ}{\mathcal{Q}}
\newcommand{\ov}{\overline}
\newcommand{\epbar}{{\ov\epsilon}{}}

\def\m#1{\mathcal#1}
\def\bea{\begin{eqnarray}}
\def\eea{\end{eqnarray}}
\def\sla{\raise.15ex\hbox{$/$}\kern-.57em}

\newcommand{\nn}{\nonumber}
\usepackage{amsmath, amsthm, amssymb,latexsym,enumerate}
\usepackage{datetime}
\newcommand{\deltab}{\boldsymbol\delta}

\newcommand{\ba}{\begin{array}}
\newcommand{\ea}{\end{array}}
\newcommand{\mc}{\mathcal}
\renewcommand{\phi}{\varphi}

\newcommand{\tabar}{\ov\theta{}}

\newcommand{\sbar}{\ov{s}{}}

\newcommand{\ep}{\epsilon}
\newcommand{\eps}{\varepsilon}
\newcommand{\da}{\delta}

\newcommand{\Qbar}{\ov{\cQ}{}}

\newcommand{\epsbar}{{\ov\varepsilon}{}}
\begin{document}

\begin{titlepage}
\vskip 2cm
\begin{center}
{\bf \Large The $\mathcal {OSP}(2,2|16)$ superconformal theory is free!}
\end{center}
\vskip 1cm

\begin{center}{
{\bf Dmitry Belyaev${}^{a}_{}$, Patrick Hearin${}^{b}_{}$, and Pierre Ramond${}^{c}_{}$
}
\vskip .9cm
{\em Institute for Fundamental Theory,\\
Department of Physics, University of Florida\\
Gainesville FL 32611, USA}
\vskip .2cm
}\end{center}

\vskip .9cm
\begin{abstract}
\noindent  
The SuperConformal theory in three space-time dimensions with $SO(16)$ $R$-symmetry,  $128$ bosons, and $128$ fermions, cannot sustain interactions.  This result is obtained using both light-cone superspace techniques which rely on algebraic consistency, and covariant methods which rely on $SO(16)$ Fierz identities which fail to produce the desired algebra. 
\end{abstract}

\vskip 1.4cm


\vspace{1cm}
\vfill \vskip 5mm \hrule width 5.cm \vskip 2mm {\small
\noindent $^a~$belyaev@phys.ufl.edu\\
\noindent $^b~$phearin@phys.ufl.edu\\
\noindent $^c~$ramond@phys.ufl.edu
}
\end{titlepage}

\section{Introduction}
According to  W. Nahm\cite{NAHM}, there are two maximally symmetric superconformal theories in three dimensions. The first  with  $\mathcal{OS}p(\,2,2\,|\,8\,)\supset SO(3,2)\times SO(8)^{}_R$ symmetry, and eight bosons and eight fermions, has been shown by Bagger, Lambert and also Gustavsson (BLG)\cite{BLG} to be a non-trivial interacting theory. 

The second, with  $\mathcal{OS}p(\,2,2\,|\,16\,)\supset SO(3,2)\times SO(16)^{}_R$ symmetry, contains $128$ bosons and $128$ fermions, is very similar to the Bagger-Lambert and Gustavsson theory. The purpose of this letter is to show that it does not sustain interactions.

We arrive at this result in two different ways. Algebraic consistency in light-cone superspace, previously  applied to the BLG theory\cite{BBKR}, shows the impossibility to construct  dynamical supersymmetry transformations which satisfy all commutation relations of this larger superalgebra. 

Covariant methods lead to the same conclusion. $SO(8)_R$ triality allows the eight bosons and eight fermions of the BLG theory to span the two $SO(8)_R$ spinor representations. The larger theory then looks similar to the BLG theory (surely at free level); all one has to do is replace $SO(8)_R$ by $SO(16)_R$, with its $128$ bosons and $128$ fermions now spanning its two spinor representations. We start with BLG-like transformations with an auxiliary vector field, but we find that the $SO(16)$ Fierz identities lead only to trivial closure of these transformations. 

\section{$N=8$ Light-Cone Superspace}
Theories with $128$ fermions and bosons are naturally described in a superspace with eight complex Grassmann variables, $\theta^m$ and $\bar\theta_m$, $m=1,\dots,8$. On the light-cone ($x^\pm_{}=(x^0_{}\pm x^3_{})/\sqrt{2},~ \partial^\pm_{}=(\partial^0_{}\pm \partial^3_{})/\sqrt{2}$), define  chiral derivatives,

\be
d^{m}_{}~=~-\frac{\partial}{\partial\bar\theta_m}\,-\,\frac{i}{\sqrt2}\,\theta^m\,\partial^+\ ;\qquad \bar d_{n}~=~ \frac{\partial}{\partial\theta^n}\,+\,\frac{i}{\sqrt2}\,\bar\theta_n\,\partial^+\ ;
\nonumber\ee
they  satisfy 

\be
\{\,d^m_{}\,,\,\bar d_{n}\,\}~=-i\,\sqrt2\,\delta^m_n\,\partial^+\ .
\ee
Introduce the constrained chiral superfield with  $256$ ($128$ bosonic, $128$ fermionic) degrees of freedom 

\bea
\Phi\,(\,y\,)\,&=&\,\frac{1}{{\p^+}^2}\,h\,(y)\,+\,i\,\theta^m\,\frac{1}{{\p^+}^2}\,{\overline \psi}_m\,(y)\,+\,i\,\theta^{mn}_{}\,\frac{1}{\p^+}\,{\overline B}_{mn}\,(y)\nn\  \\
\;&&-\,\theta^{mnp}_{}\,\frac{1}{\p^+}\,{\overline \chi}^{}_{mnp}\,(y)\,-\,\theta^{mnpq}_{}\,{\overline D}^{}_{mnpq}\,(y)+\,i\widetilde\theta^{}_{~mnp}\,\chi^{mnp}\,(y)\nn \\
&&+\,i\widetilde\theta^{}_{~mn}\,\p^+\,B^{mn}\,(y)+\,\widetilde\theta^{}_{~m}\,\p^+\,\psi^m_{}\,(y)+\,{4}\,\widetilde\theta\,{\p^+}^2\,{\bar h}\,(y)\ ,
\eea
where bar denotes complex conjugation, and 

\[
\theta^{a_1a_2...a_n}_{}~=~\frac{1}{n!}\,\theta^{a_1}\theta^{a_2}_{}\cdots\theta^{a_n}_{}\ ,\quad \widetilde\theta^{}_{~a_1a_2...a_{n}}~=~ \epsilon^{}_{a_1a_2...a_{n}b_1b_2...b_{(8-n)}}\,\theta_{}^{b_1b_2\cdots b_{(8-n)}}\,\ .
\]
The arguments of the fields are the chiral coordinates  

\[
y~=~(x,\, x^+, \,y^-\equiv x^- -\frac{i}{\sqrt2}\theta^m\bar\theta_m\, )\ ,
\]
where $x$ is the transverse variable. $\Phi$ and its complex conjugate $\overline \Phi$ satisfy the chiral constraints

\bea\label{chiralconstraints}
d_{}^m\, \Phi ~=~0, \qquad \overline d^{}_m\, \overline \Phi ~=~0\ ,
\eea
and the {\em inside-out constraint}

\begin{equation}\label{insideout}
\Phi~=~\frac{1}{4\,\p^{+4}}\,d_{}^1d_{}^2\cdots d_{}^8\, \overline\Phi.
\end{equation}
The superspace measure is given by 

\be \int d^8\theta d^8\bar\theta\,\bar \Phi\frac{1}{\partial^{+\,3}}\Phi.\ee
 The additional operators

\be
q^{m}_{}~=~-\frac{\partial}{\partial\bar\theta_m}\,+\,\frac{i}{\sqrt2}\,\theta^m\,\partial^+\ ;\quad \bar q_{n}~=~ \frac{\partial}{\partial\theta^n}\,-\,\frac{i}{\sqrt2}\,\bar\theta_n\,\partial^+\ ,
\ee
satisfy  

\be
\{\,q^m_{}\,,\,\bar q_{n}\,\}~=~i\,\sqrt2\,\delta^m_n\,\partial^+.
\ee
Since they anticommute with the chiral derivatives,  

\be
\{\,q^m_{}\,,\,\bar d^{}_n\,\}~=~\{\,q^m_{}\,,\, d^{n}_{}\,\}~=~0\ ,
\ee 
their action on the superfield do not affect its chirality. They are used to construct $SO(16)$ transformations in terms of its $SU(8)\times U(1)$ subgroup transformations, with parameters $\omega^{m}_{~~n}$, and $\omega$: 

\be\label{SO16-1}
\delta^{}_{SU(8)}\,\Phi~=~\,{i\omega^m_{~~n}}\left( q^n_{}\,\bar q_m^{}-\frac{1}{8}\delta^n_{\,m}\,q^l_{}\,\bar q_l^{}\,\right)\frac{1}{\partial^+_{}}\,\Phi\ ; 
\ee
\be\label{SO16-2}
\delta^{}_{SO(2)}\,\Phi~=~\,\frac{i\omega}{8}\left( q^m_{}\,\bar q_m^{}-\bar q_m^{}\, q^m_{}\,\right)\frac{1}{\partial^+_{}}\,\Phi.
\ee
The coset transformations, with parameters $\omega^{mn}$, and $\overline\omega_{mn}$, are given by,

\be\label{SO16-3}
\delta^{}_{\overline{coset}}\,\Phi~=~i\omega^{mn}_{}\, \bar q^{}_m\,\bar q_n^{}\,\frac{1}{\partial^+_{}}\,\Phi\ ; \quad
\delta^{}_{{coset}}\,\Phi~=~i{\omega}^{}_{mn}\,  q^{m}_{}\, q_{}^n\,\frac{1}{\partial^+_{}}\,\Phi\ . 
\ee

\section{SuperConformal Structure} 
In Dirac's light front form ($x^+=0$), the superconformal group generators split as   

$$\rm Conformal ~Group~~~\begin{cases}
~~{\rm  Lorentz~ Group: }\quad J^{+-}\ , J^+ \ ;\quad \mathcal{J}^-\\
~~{\rm  Translations: }\quad P\ , P^+\ ;\quad \mathcal{P}^-\\
~~{\rm  Dilatation: }\quad D \\
~~{\rm  Conformal: }\quad  K\ ,  K^+\ ;\quad \mathcal{K}^-
\end{cases}\ ,$$
 $$~~~~~~~~~~~~\rm Supers~~~\begin{cases}
~~{\rm  Supersymmetry: }\quad q\ , \bar q\ ;\quad {\mathcal Q}\ , \overline{\mathcal Q}\\
~~{\rm  Superconformal: }\quad s\ , \bar s\ ;\quad {\mathcal S}\ , \overline{\mathcal S}
\end{cases}\ ,$$
with the dynamical generators written in capital calligraphic letters. Note that $J^+$ and $K^+$  at $x^+=0$, and $R$-symmetries (Eqs.(\ref{SO16-1}-\ref{SO16-3})) are all kinematical.

With superconformal symmetries,  {\em all} dynamical operators are obtained by commutation of the dynamical supersymmetry transformations with  kinematical operators, so that  dynamical supersymmetry transformations determine the full theory.

\subsection{Kinematical Transformations} 
Superconformal kinematical transformations are linear in the fields, even in the interacting theory, starting with  

\be
\da^{}_{P^{+}}\,\Phi^a_{}~=-i\,\p^+_{}\,\Phi^a_{}\ ;\qquad 
\da^{}_{P}\,\Phi^a_{}~=-i\,\p\,\Phi^a_{}\ ; \nn
\ee
\be
\da^{}_{J^{+}}\,\Phi^a_{}~=~i x\,\p^+_{}\,\Phi^a_{}\ ;\qquad 
\da^{}_{J^{+-}}\,\Phi^a_{}~=~i(\, \mc{A}+\frac{x}{2}\p+2 \,)\,\Phi^a_{}\ ; \nn
\ee
\be
\da^{}_{D}\,\Phi^a_{}~=~i\,(\,\mc A\,-\,\frac{x}{2}\,\p\,)\,\Phi^a_{}\ ;\quad
\da^{}_{K}\,\Phi^a_{}~=~2i\,x\,\mc A\,\Phi^a_{} \ ;\quad
\da^{}_{K^+}\,\Phi^a_{}~=i\,x^2\p^+_{}\,\Phi^a_{}\ . 
\label{kinconf}
\ee
where $\p$ is the transverse derivative, and
\be
\label{defA}
{\mc A}~\equiv~x^-\,\p^+-\frac{x}{2}\,\p\,-\frac{1}{2}{\mc N}+\frac{3}{2}\ ;\qquad{}
{\mc N}~\equiv~\sum_{m=1}^{8}\left(\theta^m_{}\frac{\p}{\p\theta^m_{}}\,+\,\tabar^{}_m\frac{\p}{\p\tabar^{}_m}\right)\ .
\ee
The taxonomic index $a$ allows for several superfields. The kinematical (spectrum generating) supersymmetries are  
\be
\da^{\,}_{\ep\qbar}\,\Phi^a_{}~=~\ep^m_{}\qbar_m^{}\,\Phi^a_{}\ ; \qquad \da^{\,}_{\epbar q}\,\Phi^a_{}~=~\epbar^{}_m q^m_{}\,\Phi^a_{}\ ,
\ee
and the kinematical superconformal transformations are
\be
\da^{}_{\ep\sbar}\,\Phi^a_{}~=-i x\, \ep^m\,\qbar_m^{}\,\Phi^a_{}\ ;\qquad
\da^{}_{\epbar s}\,\Phi^a_{}~=~i x\, \epbar_m\,q^m_{}\,\Phi^a_{}\ .
\ee
where $\ep^m$ and $\epbar_m$ are eight anticommuting parameters.

\subsection{Dynamical Transformations}
The free dynamical supersymmetry transformations (in boldface) are,  

\be
\label{freeds}
\dab^{free}_{\ep\Qbar}\,\Phi^a~=~\frac{1}{\sqrt{2}}\ep_{}^m\qbar^{}_m\,\frac{\p}{\p^+_{}}\,\Phi^a_{}\ ,\qquad 
\dab^{free}_{\epbar\cQ}\,\Phi^a~=~\frac{1}{\sqrt{2}}\epbar^{}_m q^{m}_{}\,\frac{\p}{\p^+_{}}\,\Phi^a_{}\ .\ee
It is easy to see that they are chiral and satisfy the inside-out constraint. 

In superconformal theories, {\em all} dynamical information is contained in the dynamical supersymmetry transformations; knowing them suffices to generate all interactions. We set

\bea
\dab_{\ep\Qbar}\Phi^a=\dab_{\ep\Qbar}^{free}\Phi^a+\dab_{\ep\Qbar}^{int}\Phi^a, \quad
\dab_{\epbar\cQ}\Phi^a=\dab_{\epbar\cQ}^{free}\Phi^a+\dab_{\epbar\cQ}^{int}\Phi^a\ ,
\eea
and proceed to determine the form of $\dab_{\epbar\cQ}^{int}\Phi^a$ and its conjugate. Our procedure parallels \cite{BBKR}, except for the number of Grassmann variables.
\vskip .3cm

\noindent The interacting parts must satisfy chirality, and the inside out constraint

\be
\label{CC}
d^m\left( \dab_{\ep\Qbar}^{int}\Phi^a \right)=0, \qquad \dab_{\epbar\cQ}^{int}\Phi^a=\frac{d^8}{4\p^{+4}}\left( \dab_{\ep\Qbar}^{int}\Phi^a \right)^\ast
\ee

\noindent Additional requirements from kinematics are: 

\begin{enumerate}[(i)]

\item \label{2}  Independence from  $x^-$ and $x$, since

\be
[\,\delta^{}_{P^+}\,,\,\deltab^{}_{\epsilon\overline {\mathcal Q}}\,]\,\Phi^a~=~[\,\delta^{}_{P^+}\,,\,\deltab^{}_{\overline\epsilon {\mathcal Q}}\,]\,\Phi^a~=~0
\ ,\ee
and 

\be
[\,\delta^{}_{P}\,,\,\deltab^{}_{\epsilon\overline {\mathcal Q}}\,]\,\Phi^a~=~[\,\delta^{}_{P}\,,\,\deltab^{}_{\overline\epsilon {\mathcal Q}}\,]\,\Phi^a~=~0\ .
\ee

\item \label{4} No transverse derivatives $\partial$, since 

\be
[\,\delta^{}_{J^+}\,,\,\deltab^{}_{\overline\epsilon {\mathcal Q}}\,]\,\Phi^a~=~\frac{i}{\sqrt{2}}\,\delta^{}_{\bar\epsilon q}\,\Phi^a\ ,\quad [\,\delta^{}_{J^+}\,,\,\deltab^{}_{\epsilon\overline {\mathcal Q}}\,]\,\Phi^a~=~\frac{i}{\sqrt{2}}\,\delta^{}_{\epsilon\bar q}\,\Phi^a\ ,
\ee
lead to

\be
[\,\delta^{}_{J^+}\,,\,\deltab^{int}_{\overline\epsilon {\mathcal Q}}\,]\,\Phi^a~=~[\,\delta^{}_{J^+}\,,\,\deltab^{int}_{\epsilon\overline {\mathcal Q}}\,]\,\Phi^a~=~0\ .\ee

\item \label{5} From 

\be
[\,\delta^{}_{\bar\epsilon q}\,,\,\deltab^{}_{\epsilon\overline {\mathcal Q}}\,]\,\Phi^a~=-\bar\epsilon_m\epsilon^m\,\delta^{}_P\,\Phi^a_{}\ ,\qquad [\,\delta^{}_{\epsilon \bar q}\,,\,\deltab^{}_{\overline\epsilon {\mathcal Q}}\,]\,\Phi^a~=~\bar\epsilon_m\epsilon^m\,\delta^{}_P\,\Phi^a_{}
\ ,\ee
we deduce, 

\be 
[\,\delta^{}_{\bar\epsilon q}\,,\,\deltab^{int}_{\epsilon\overline {\mathcal Q}}\,]\,\Phi^a~=~[\,\delta^{}_{\epsilon \bar q}\,,\,\deltab^{int}_{\overline\epsilon {\mathcal Q}}\,]\,\Phi^a~=~0\ .
\ee

\item \label{6} Proper transformations under $J^{+-}$ and $D$ require

\be
[\,\delta^{}_{J^{+-}}\,,\,\deltab^{int}_{\epsilon\overline {\mathcal Q}}\,]\,\Phi^a~=~\frac{i}{2}\,\deltab^{int}_{\epsilon\overline {\mathcal Q}}\,\Phi^a\ ,
\qquad [\,\delta^{}_{J^{+-}}\,,\,\deltab^{int}_{\overline\epsilon {\mathcal Q}}\,]\,\Phi^a~=~\frac{i}{2}\,\deltab^{int}_{\overline\epsilon {\mathcal Q}}\,\Phi^a.
\ee

\be
[\,\delta^{}_{D}\,,\,\deltab^{int}_{\epsilon\overline {\mathcal Q}}\,]\,\Phi^a~=-\frac{i}{2}\,\deltab^{int}_{\epsilon\overline {\mathcal Q}}\,\Phi^a\ ,\qquad [\,\delta^{}_{D}\,,\,\deltab^{int}_{\overline\epsilon {\mathcal Q}}\,]\,\Phi^a~=-\frac{i}{2}\,\deltab^{int}_{\overline\epsilon {\mathcal Q}}\,\Phi^a\ .
\ee

\item \label{8} The correct $U(1)$ $R$-charge,

\be
[\,\delta^{}_{U(1)}\,,\,\deltab^{int}_{\epsilon\overline {\mathcal Q}}\,]\,\Phi^a~=-\frac{1}{2}\,\deltab^{int}_{\varepsilon\overline {\mathcal Q}}\,\Phi^a\ ,\qquad [\,\delta^{}_{U(1)}\,,\,\deltab^{int}_{\overline\epsilon {\mathcal Q}}\,]\,\Phi^a~=~\frac{1}{2}\,\deltab^{int}_{\overline\epsilon {\mathcal Q}}\,\Phi^a\ .
\ee

\item \label{9} The sixteen interacting supersymmetries must transform as an $SO(16)$ vector, which, in the $SU(8)\times U(1)$ decomposition, means,  

\be\label{cosetbar}
[\,\delta^{}_{\overline{coset}}\,,\,\deltab^{}_{\varepsilon\overline {\mathcal Q}}\,]\,\Phi^a~=~0\ , 
\qquad 
[\,\delta^{}_{{coset}}\,,\,\deltab^{}_{\varepsilon\overline {\mathcal Q}}\,]\,\Phi^a~=~\deltab^{}_{ \bar\varepsilon'{\mathcal Q}}\,\Phi^a\ ,
\ee
with $\bar\epsilon'^{}_m=2\overline\omega^{}_{mn}\epsilon^n_{}$. Similarly, with $\epsilon'^m=2\omega^{mn}\overline\epsilon_n$,

\be\label{cosetbar2}
[\,\delta^{}_{\overline{coset}}\,,\,\deltab^{}_{\overline\epsilon {\mathcal Q}}\,]\,\Phi^a~=~\deltab^{}_{ \epsilon'\overline{\mathcal Q}}\,\Phi^a\ , 
\qquad 
[\,\delta^{}_{{coset}}\,,\,\deltab^{}_{\overline\epsilon {\mathcal Q}}\,]\,\Phi^a~=~0\ .
\ee

\item \label{10} Dynamical interacting supersymmetries are cubic powers of superfields. 

In three dimensions, Bose fields have mass dimension one-half, so $\Phi$ has half-odd integer dimension, assuming integer power of $\p^+$.  In a conformal theory with no dimensionful parameters, the interacting supersymmetry must then be odd powers of superfields. Define

\be
\da^{}_\Delta\,\Phi^a~\equiv~\da^{}_{(J^{+-}-D)}\,\Phi^a~=~i\left(x\partial+\frac12\right)\,\Phi^a\ .\ee
Since $\dab_{\ep\Qbar}^{int}\Phi^a$  contain no transverse variable, it follows that 

\bea
[\da^{}_\Delta,\dab_{\ep\Qbar}^{int}]\Phi^a=\frac{i}{2}(n_\Phi-1)\dab_{\ep\Qbar}^{int}\Phi^a\ ,
\eea
where $n_\Phi$ is the number of superfields. On the other hand, by matching with the free part, the algebra requires 
  
\bea
\label{kcDelta}
[\da^{}_\Delta,\dab_{\ep\Qbar}^{int}]\Phi^a=i\dab_{\ep\Qbar}^{int}\Phi^a,
\eea
which is consistent for $n_\Phi=3$. This theory must contain a tensor with at least four indices, $f^{abcd}$, as in the BLG theory.

\end{enumerate}
\vskip .5cm

The dynamical supersymmetry transformations are written in terms of the basic cubic nested form,

\bea
K^{}_\alpha &\equiv &\frac{1}{\partial^{+A_\alpha}}\left((\partial^{+B_{\alpha}}\,\varphi^b_{}\,)
\, \frac{1}{\partial^{+M_\alpha}}\left(\,\partial^{+C_\alpha}\,\varphi^c\,)(\,\partial^{+D_\alpha}\,\varphi^d\,)\,\right)\right)\\ 
&\equiv& \frac{1}{\partial^{+A_\alpha}}\left(\partial^{+B_{\alpha}}\,,
\, \frac{1}{\partial^{+M_\alpha}}\left(\,\partial^{+C_\alpha}\,,\,\partial^{+D_\alpha}\,\right)\right),
\eea
and the coherent state operators\cite{BRINK},

\be 
E_{\eta}^{}~=~e^{\,\eta\cdot\widehat{\overline d}}_{},\quad E^{}_r=~e^{\,r\hat\partial}_{},\quad E_{\varepsilon}^{}~=~e^{\,\varepsilon\cdot\widehat{\overline q}}_{}\ ,\qquad  E_{\,\bar\varepsilon}^{}~=~e^{\,\bar\varepsilon\cdot{\widehat q}}.
\ee
The Grassmann variables $\eta^m,\, \zeta^m$, and $r,\,r'$ are dummy variables, and $\varepsilon^m\,(\ov\varepsilon_m)$ are the supersymmetry parameters. 

Using insertion  operators ${\mc U}_i$, $i=1,2,3,4$, we define

\bea
K^{(\eta,\zeta)}_\alpha&\equiv&\frac{1}{\partial^{+A_\alpha}}\left(E_{\eta}^{}\partial^{+B_{\alpha}}\,,
\, E_{-\eta}^{}\frac{1}{\partial^{+M_\alpha}}\left(\,E_{\zeta}^{}\partial^{+C_\alpha}\,,E_{-\zeta}^{}\,\partial^{+D_\alpha}\,\right)\right)\\
&\equiv&\left(E_{\eta}^{}{\mc U}_1\right)\left(E_{-\eta}^{}{\mc U}_2\right)\left(E_{\zeta}^{}{\mc U}_3\right)\left(E_{-\zeta}^{}{\mc U}_4\right)K^{}_\alpha
\eea
and
\be
K^{(r,r')}_\alpha=\frac{1}{\partial^{+A_\alpha}}\left(E_{r}^{}\partial^{+B_{\alpha}}\,,
\, E_{-r}^{}\frac{1}{\partial^{+M_\alpha}}\left(\,E_{r'}^{}\partial^{+C_\alpha}\,,E_{-r'}^{}\,\partial^{+D_\alpha}\,\right)\right)
\ee
Supersymmetry parameters are introduced through,

\bea
K^{(\varepsilon;\eta,\zeta)}_\alpha&\equiv&\left(E_{\varepsilon}^{}{\mc U}_1\right)\left(E_{-\varepsilon}^{}{\mc U}_2\right)K^{(\eta,\zeta)}_\alpha,\\
&=&
\frac{1}{\partial^{+A_\alpha}}\left(E_{\varepsilon}^{}E_{\eta}^{}\partial^{+B_{\alpha}}\,,
\, E_{-\varepsilon}^{}E_{-\eta}^{}\frac{1}{\partial^{+M_\alpha}}\left(\,E_{\zeta}^{}\partial^{+C_\alpha}\,,E_{-\zeta}^{}\,\partial^{+D_\alpha}\,\right)\right).
\eea
In this notation, the dynamical supersymmetry transformations (showing explicitly the superfield taxonomic indices) are of the form, 

\be
\delta^{int}\Phi^a= f^a_{\ bcd}\sum_\alpha {K}^{(\varepsilon;\eta,\zeta)bcd}_\alpha.
\ee
They are similar to those in the BLG theory\cite{BBKR}, except that the indices now run over eight values. Some new features must be noted: 

\begin{itemize}

\item The correct $U(1)$ charge now requires

\bea(\eta^m\frac{\partial}{\partial\eta^m}+\zeta^m\frac{\partial}{\partial\zeta^m}-8)K^{(\epsilon,\eta,\zeta)a}_\alpha=0,\eea
so that only terms octal in $\eta,\zeta$ need to be considered, that is

\be \zeta^8,\ \eta\zeta^7,\  \eta^2\zeta^6,\ \eta^3\zeta^5,\ \eta^4\zeta^4,\ \eta^5\zeta^3,\ \eta^6\zeta^2,\ \eta^7\zeta,\ \eta^8.
\ee

\item The proper transformation under $J^{+-}$, together with the $U(1)$ constraint restricts the number of $\p^+$ derivatives to eight,

\be -A_\alpha+B_\alpha-M_\alpha+C_\alpha+D_\alpha=8.
\ee

\item The correct transformation properties under the coset transformations splits the Ansatz into two types, the even Ansatz where the sum is over ($\eta^8,\eta^6\zeta^2,\eta^4\zeta^4,\eta^2\zeta^6,\zeta^8$), and odd Ansatz over ($\eta^7\zeta,\eta^5\zeta^3,\eta^3\zeta^5,\eta\zeta^7$), with recursion relations

\begin{align}&A_\alpha=A_{-\alpha}-2\alpha,&& B_\alpha=B_{-\alpha}-2\alpha,&& M_\alpha=M_{-\alpha}+4\alpha,\nonumber \\ \nonumber \\
&C_\alpha=C_{-\alpha}+2\alpha,&& D_\alpha=D_{-\alpha}+2\alpha.\end{align}

\end{itemize}

These constraints further narrow the form of the dynamical supersymmetries. In terms of

\be
K^{}_{\alpha(k,8-k)}=\frac{\epsilon^{i_1\dots i_{8}}}{k!(8-k)!}\frac{\partial}{\partial\eta^{i_1\dots i_k}}\frac{\partial}{\partial\zeta^{i_{k+1}\dots i_{8-k}}}K^{(\eta,\zeta)}_\alpha
\ee
evaluated at $\eta=\zeta=0$, we find two linear combinations which satisfy chirality, as well as all kinematical and inside-out constraints, the odd Ansatz, 

\be
\delta^{int}_{odd}\Phi^a=\sum_{odd}K^a_{\alpha}= f^a_{\ bcd}\sum_{\alpha=-\frac{3}{2}}^{\frac{3}{2}}(-1)^{\alpha+\frac{1}{2}}_{}K^{bcd}_{\alpha(4-2\alpha,4+2\alpha)},
\ee
and the even Ansatz,

\be
\delta^{int}_{even}\Phi^a=\sum_{even}K^a_{\alpha}= f^a_{\ bcd}\sum_{\alpha=-2}^{2}(-1)^\alpha_{}K^{bcd}_{\alpha(4-2\alpha,4+2\alpha)}.
\ee

All kinematic requirements being satisfied, we use these expressions to find the other dynamical transformations. The light-cone Hamiltonian is calculated from,

\bea
[\dab_{\eps\Qbar}^{free}+\dab_{\eps\Qbar}^{int},\dab_{\epsbar\cQ}^{free}+\dab_{\epsbar\cQ}^{int}]\phi^a
=\sqrt2\epbar_m\eps^m\dab_{\mc{P}^{-}}\phi^a.\eea
The boost $\delta_{\m J^-}\Phi^a$ are obtained from the commutator of the Hamiltonian with the transverse conformal transformation,

\be
2i\delta^{}_{\mc J^-}\Phi^a~=~[\,\delta^{}_K\,,\,\delta^{}_{\mc P^-}\,]\Phi^a.\ee
 
The first consistency check comes from the commutator of the hamiltonian with the boost which, by conformal symmetry, should vanish,

\be
[\,\delta^{}_{\mc P^-}\,,\,\delta^{}_{\mc J^-}\,]\Phi^a~=~0\ .\ee
We now show that, unlike in the BLG case, it does not vanish. For the odd Ansatz (similar conclusions apply as well to the even Ansatz), and to first order in $f^{abcd}$, a long calculation shows this commutator to be proportional to 

\bea
&&\sum_{odd}^{}{\mc A}_\alpha\left(K^{(rr',1)}_\alpha-K^{(1,rr')}_{\alpha+1}\right)+2\sum_{even}{\mc A}_\alpha K^{(r,r')}_{\alpha+\frac{1}{2}}-\nonumber\\
&&~~~-\sum_{even}{\mc B}_\alpha\left(K^{(rr',1)}_{\alpha+\frac{1}{2}}-K^{(1,rr')}_{\alpha+\frac{3}{2}}\right)-2\sum_{odd}{\mc B}_\alpha K^{(r,r')}_{\alpha+1}
\eea
where

\bea
{\mc A}_\alpha&=&(B_\alpha^{}+\alpha-\frac{7}{2})\partial^+{\mc U}_2+(M^{}_\alpha-C^{}_\alpha-D^{}_\alpha+5)\partial^+{\mc U}_1,\\
{\mc B}_\alpha&=&\left((C^{}_\alpha-\alpha-4)\frac{1}{\partial^+}{\mc U}_3-(D^{}_\alpha-\alpha-4)\frac{1}{\partial^+}{\mc U}_4\right)(\partial^+{\mc U}_1)(\partial^+{\mc U}_2).
\eea
In these expressions, the coefficients $(A_\alpha,B_\alpha,M_\alpha,C_\alpha,D_\alpha)$ have yet to be determined. 

The choices $(A_{-\frac{3}{2}},B_{-\frac{3}{2}},M_{-\frac{3}{2}},C_{-\frac{3}{2}},D_{-\frac{3}{2}})=(5,5,-6,1,1)$ eliminate the largest number of terms, and we focus on those the remaining terms which contain the combination,

\be
Z=\partial^2\frac{1}{8!}\epsilon^{i_1\dots i_8}_{}\ov d_{i1}\dots \ov d_{i_8}.\ee
Their contributions to the commutator reduce to,  

\be\label{com}
\left[\hat {\mc U}_3\,{\mc Z}_1-\hat {\mc U}_4\,\mathfrak{g}_1\,{\mc Z}_3+(\hat {\mc U}_3-\hat {\mc U}_4)(\mathfrak{g}_2-\mathfrak{t} \mathfrak{g}_1){\mc Z}_3\frac{1}{\partial^+}(\partial^+{\mc U}_1)(\partial^+ {\mc U}_2) \right]K^{}_{-\frac{3}{2}},
\ee 
where 

\be
(\frac{1}{\partial^+}{\mc U}_3)\equiv\hat {\mc U}_i, \quad Z{\mc U}_i\equiv {\m Z}_i;\qquad \mathfrak{s}\equiv (\frac{1}{\partial^+}{\mc U}_2)(\partial^+{\mc U}_3), \quad \mathfrak{t}\equiv (\frac{1}{\partial^+}{\mc U}_1)(\partial^+{\mc U}_4),
\ee
with

\bea
\mathfrak{g}_1&=&32\mathfrak{s}^7(1+\mathfrak{t})^2-32\mathfrak{s}^6(1+\mathfrak{t})^3+6\mathfrak{s}^5(1+\mathfrak{t})^4,\\
\mathfrak{g}_2&=&32\mathfrak{s}^8(1+\mathfrak{t})^2-48\mathfrak{s}^7(1+\mathfrak{t})^3+18\mathfrak{s}^6(1+\mathfrak{t})^4-\mathfrak{s}^5(1+\mathfrak{t})^5.
\eea
Further simplifications reduce Eq.(\ref{com}) to,

\bea
&&\frac{1}{\partial^{+6}}\left(\,\partial^{+8}X,\left[ 32\partial^{+2}(\partial^{+4},~)-48\partial^{+3}(\partial^{+3},~)+\right.\right.\nonumber\\
&& \left.\left. ~~~~~~~+18\partial^{+4}(\partial^{+2},~)-\partial^{+5}(\partial^{+},~)\right]\right)+{\mc O}(\partial^{+7}X).
\eea
Since these clearly do not vanish, we conclude that an interacting theory based on $OSp(2,2|16)$ symmetry does not exist. In the next section, we arrive at the same conclusions using the familiar covariant description.

\section{Covariant Formulation}
The covariant form of the BLG theory begins with the supersymmetry transformations,

\bea
\delta\,X^{I}_{\,a}&=&i\bar\epsilon\Gamma^I_{}\Psi_a^{} \nonumber\\
\delta\, \Psi_{a}^{}&=& {\mc D}^{}_\mu X^{I}_{a}\gamma^\mu_{}\Gamma^I_{}\epsilon-\frac{1}{6}X^{I}_bX^{J}_{c}X^{K}_{d}f^{bcd}_{~~~\,a}\Gamma^{IJK}_{}\epsilon\nonumber\\
\delta \tilde{A}_{\mu~\,a}^{~\,b}&=&i\bar\epsilon\gamma^{}_\mu\Gamma^I_{}X^I_{c}\Psi^{}_df^{cdb}_{~~~\,a},
\eea
where $I$ is the $SO(8)$ vector index. The supersymmetry parameters, $\epsilon_{\alpha A}$, span one $SO(8)$ spinor representation, where $\alpha=0,1,2$ and $A=1,2\dots, 8$.   The fermions $\Psi_{\alpha\dot A\,a}$, where $\dot A=1,\dots,8$, span the other $SO(8)$ spinor representation. The Dirac matrices are $(\Gamma^I)^{}_{A\dot A}$, and  $\gamma^{}_{\alpha}$, with $\gamma^{012}\epsilon=\epsilon$. The covariant derivatives are given by

\be
{\mathcal D}^{}_\mu X^I_a~=~\partial^{}_\mu X^I_a+\tilde A_{\mu~\,a}^{~\,b}X^I_b\ .\qquad {\mathcal D}^{}_\mu \Psi^{}_{\dot A\,a}~=~\partial^{}_\mu \Psi^{}_{\dot A\,a}+\tilde A_{\mu~\,a}^{~\,b}\Psi^{}_{\dot A\,b},
\ee
where the vector field $\tilde A_{\mu~\,a}^{~\,b}$, with the dimension of mass, is an auxiliary field (canonical boson fields have mass dimension one-half). Two supersymmetry transformations generate, as expected, a translation and a gauge transformation,

\be
[\,\delta_1\,,\,\delta_2\,]X^I_a~=~v^\mu_{}\partial_\mu^{}X^I_a+\Lambda^a_{~b}X^I_b,
\ee
where the translation and gauge parameters are 

\be
 v^{}_\mu=\ov\epsilon^{}_2\gamma^{}_\mu\epsilon^{}_1,\qquad \Lambda^a_{~b}=v^{}_\mu \widetilde A^a_{\mu\,b}+f^{a}_{~bcd}(\ov\epsilon^{}_{2}\Gamma^{IJ}\epsilon^{}_{1})X^I_cX^J_d.
 \ee
This equation, which relies heavily on the $SO(8)$ Fierz identities, shows the dependence of  the gauge parameters on the bosonic coordinates.

The possible generalization of this algebra to $SO(16)$ begins with an alternative description based on $SO(8)$ triality\cite{NISHINO}. The vector index $I$ is simply replaced by a spinor index $A$; the bosons are labelled by $X_{A\,_a}$, and the supersymmetry parameters transform as $SO(8)$ vectors,  $\epsilon^I_\alpha$. This leads to the new transformation rules,   

\bea
\delta\,X^{}_{A\,a}&=&\bar\epsilon^I_{}(\Gamma^I)_{A\dot A}\Psi^{}_{\dot A\,a} \nonumber\\
\delta\, \Psi_{\dot Aa}^{}&=& {\mathcal D}^{}_\mu X^{}_{A\,a}\gamma^\mu_{}(\Gamma^I)^{}_{\dot A\,A}\epsilon^I_{}+\frac{1}{6}f^{bcd}_{~~~\,a}X^{}_{A\,b}X^{}_{B\,c}X^{}_{C\,d}(\Gamma^{J})^{}_{\dot A\,A}(\Gamma^{IJ}_{})^{}_{BC}\epsilon^I_{}\nonumber\\
\delta \tilde{A}_{\mu~\,a}^{~\,b}&=&f^{cdb}_{~~~\,a}\bar\epsilon^I_{}\gamma^{}_\mu\Psi^{}_{\dot A\,d}(\Gamma^I)^{}_{\dot A\,B}X^{}_{B\,c}.
\eea
As expected, these commutators have the BLG structure, and close on translations and gauge transformations\cite{NISHINO}.
\vskip .3cm

We seek a generalization of this algebra where the $128$ bosonic coordinates span one $SO(16)$ spinor representation, and the fermionic coordinates the other. We label the first spinor by $A=1,\dots,128$, the second by  $\dot A=1,\dots,128$. The vector index $I$ runs over sixteen values.

We posit the supersymmetry transformations of the bosons,

\be
\delta\,X^{}_{A\,a}=\bar\epsilon^I_{}(\Gamma^I)_{A\dot A}\Psi^{}_{\dot A\,a}.
\ee 
The fermions' supersymmetry transformations contain the same covariant derivatives with the auxiliary one-form,

 \be
 \delta\, \Psi_{\dot Aa}^{}= {\mathcal D}^{}_\mu X^{}_{A\,a}\gamma^\mu_{}(\Gamma^I)^{}_{\dot A\,A}\epsilon^I_{}+\delta' \Psi_{\dot Aa}^{},
 \ee
augmented by $\delta'$, the most general transformation allowed by $SO(16)$.

This supersymmetry variation transforms as $\bf{1920}$ vector-spinor representation, and from,  

\be
({\bf 128}\times {\bf 128}\times{\bf 128})_a~=~\bf{326144}+\bf{13312}+\bf{1920},
\ee
we see that $\delta' \Psi_{\dot Aa}^{}$  contains cubic antisymmetric products of bosons. 

Bi-spinors transform as antisymmetric forms,
\be
{\bf 128}\times {\bf 128}~=~[{\bf 1}+{\bf 1820}+{\bf 6435}]_{sym}+[{\bf 8008}+{\bf 120}]_{anti}
\ee
where ${\bf 6435}$ is the self-dual eight form. Only the two- and six-forms appear in their antisymmetric product.   
For any two spinors, $Y_A$ and $Z_B$, we thus have 

\bea
128Y^{}_AZ^{}_B&=&(1)^{}_{AB}(YZ)-\frac{1}{2}(\Gamma^{(2)}_{})^{}_{AB}(Y\Gamma^{(2)}_{}Z)+\frac{1}{4!}(\Gamma^{(4)}_{})^{}_{AB}(Y\Gamma^{(4)}_{}Z)\nonumber\\
&&-\frac{1}{6!}(\Gamma^{(6)}_{})^{}_{AB}(Y\Gamma^{(6)}_{}Z)+\frac{1}{2\cdot8!}(\Gamma^{(8)}_{})^{}_{AB}(Y\Gamma^{(8)}_{}Z),
\eea
from which all Fierz identities are derived. The antisymmetric product of two bosons comes in two covariant expressions,

\be
X^{}_{Ab}(\Gamma^{I_1I_2I_3I_4I_5I_6}_{})_{AB}\,X^{}_{Bc}\equiv (X_{b}\Gamma^{(6)}X_c)\ ,\quad X^{}_{Ab}(\Gamma^{I_1I_2}_{})_{AB}\,X^{}_{Bc}\equiv (X_{b}\Gamma^{(2)}X_c).
\ee
In contrast, $SO(8)$ has  only one antisymmetric covariant expression, $X_b\Gamma^{(2)}X_c$. Hence the most general form of the supersymmetry variation is

\bea
\delta' \Psi_{\dot Aa}^{}&=&f_a^{~bcd}\epsilon^I_{}\left[ k_1(\Gamma^J_{}X^{}_b)(X^{}_c\Gamma^{IJ}_{}X^{}_d)+k_2(\Gamma^{I(2)}_{}X^{}_b)(X^{}_c\Gamma^{(2)}_{}X^{}_d)\right.\nonumber\\
&&~~~~\left.+ k_3(\Gamma^{(5)}_{}X^{}_b)(X^{}_c\Gamma^{I(5)}_{}X^{}_d)+k_4(\Gamma^{I(6)}_{}X^{}_b)(X^{}_c\Gamma^{(6)}_{}X^{}_d)\right]
\eea
Assuming antisymmetry of $f^{abcd}$, direct evaluation of the commutator of two supersymmetries yields,  

\bea
[\,\delta_1^{}\,,\,\delta^{}_2\,]\,X^{}_{A\,a}&=&(\ov \epsilon_2^I\gamma^\mu_{}\epsilon_1^I){\mc D}^{}_\mu X^{}_{A\,a}+\nonumber\\
&&
+f_a^{~bcd}\epsilon^{IJ}_{}\left[ k_1(\Gamma^I_{}\Gamma^K_{}X^{}_b)_A^{}(X^{}_c\Gamma^{JK}_{}X^{}_d)+k_2(\Gamma^I_{}\Gamma^{J(2)}_{}X^{}_b)_A^{}(X^{}_c\Gamma^{(2)}_{}X^{}_d)+\right.\nonumber\\
&&~~~~~~~~~~~~~~~\left.+ k_3(\Gamma^I_{}\Gamma^{(5)}_{}X^{}_b)_A(X^{}_c\Gamma^{J(5)}_{}X^{}_d)+k_4(\Gamma^I_{}\Gamma^{J(6)}_{}X^{}_b)^{}_A(X^{}_c\Gamma^{(6)}_{}X^{}_d)\right],\nonumber\\
&&\eea
where $
\epsilon^{IJ}_{}=\ov\epsilon_2^I\epsilon_1^J-(I\leftrightarrow J).$ We use,

\be
\Gamma^I_{}\Gamma^K_{}=\Gamma^{IK}_{}+\delta^{IK}_{},\ee
and apply the Fierz transformations to obtain

\be
(\Gamma^{[I(5)}X^{}_b)^{}_A(X^{}_c\Gamma^{J](5)}_{}X^{}_d)=\sum_{n=0,2,4}c^{}_n(\Gamma^{(2n)}X^{}_b)^{}_A
(X^{}_c\Gamma^{[J(5)}_{}\Gamma^{(2n)}_{}\Gamma^{I](5)}_{}X^{}_d),
\ee

\be
(\Gamma^{(4)}X^{}_b)^{}_A(X^{}_c\Gamma^{IJ(4)}_{}X^{}_d)=\sum_{n=0,2,4}c^{}_n(\Gamma^{(2n)}X^{}_b)^{}_A
(X^{}_c\{\Gamma^{IJ(4)}_{}\Gamma^{(2n)}_{}\Gamma^{(4)}+\Gamma^{(4)}\Gamma^{(2n)}_{}\Gamma^{IJ(4)}_{}\}X^{}_d)
\ee
and for $k=2,6$, 

\be
(\Gamma^{IJ(k)}X^{}_b)^{}_A(X^{}_c\Gamma^{(k)}_{}X^{}_d)=\sum_{n=0,2,4}c^{}_n(\Gamma^{(2n)}X^{}_b)^{}_A
(X^{}_c\{\Gamma^{IJ(k)}_{}\Gamma^{(2n)}_{}\Gamma^{(k)}+\Gamma^{(k)}\Gamma^{(2n)}_{}\Gamma^{IJ(k)}_{}\}X^{}_d),
\ee
where $128(c^{}_0,c^{}_2,c^{}_4)=(1,\frac{1}{2}, \frac{1}{2\cdot 4!})$ are Fierz coefficients. Use of the identities 

\bea
\Gamma^{(k)}_{}\Gamma^{(n)}_{}\Gamma^{(k)}_{}&=&g^{}_{n,k}\Gamma^{(n)}_{},\\
\Gamma^{[I(k)}_{}\Gamma^{(n)}_{}\Gamma^{J](k)}_{}&=&  a^{}_{n,k}\Gamma^{[I}_{}\Gamma^{(n)}\Gamma^{J]}_{}+b^{}_{n,k}\{
\Gamma^{IJ}_{}\Gamma^{(n)}_{}+\Gamma^{(n)}_{}\Gamma^{IJ}_{}\},\nonumber\\
&&\\
\Gamma^{[IJ(k)}_{}\Gamma^{(n)}_{}\Gamma^{(k)}_{}+\Gamma^{(k)}_{}\Gamma^{(n)}_{}\Gamma^{[IJ(k)}_{}&=&  c^{}_{n,k}\Gamma^{[I}_{}\Gamma^{(n)}\Gamma^{J]}_{}+d^{}_{n,k}\{
\Gamma^{IJ}_{}\Gamma^{(n)}_{}+\Gamma^{(n)}_{}\Gamma^{IJ}_{}\}.\nonumber\\
\eea
expresses the commutator in terms of four combinations,

\bea
A^{IJ}_{a}&\equiv&f^{~bcd}_aX^{}_{bA}(X^{}_c\Gamma^{IJ}_{}X^{}_d),\nonumber\\
B^{IJ}_{a}&\equiv&f^{~bcd}_a(\Gamma^{(4)}X^{}_b)^{}_A(X^{}_c\Gamma^{[I}_{}\Gamma^{(4)}_{}\Gamma^{J]}_{}X^{}_d),\nonumber\\
C^{IJ}_{a}&\equiv&f^{~bcd}_a(\Gamma^{(4)}X^{}_b)^{}_A(X^{}_c\{\Gamma^{IJ}_{}\Gamma^{(4)}_{}+\Gamma^{(4)}_{}\Gamma^{IJ}_{}\}X^{}_d),\nonumber\\
D^{IJ}_{a}&\equiv&f^{~bcd}_a(\Gamma^{(8)}X^{}_b)^{}_A(X^{}_c\{\Gamma^{IJ}_{}\Gamma^{(8)}_{}+\Gamma^{(8)}_{}\Gamma^{IJ}_{}\}X^{}_d).
\eea
Not all are independent as they satisfy two equations,

\bea
c^{}_2(a^{}_{4,3}A^{IJ}_{a}+b^{}_{4,3}B^{IJ}_{a})+c^{}_4b^{}_{8,3}C^{IJ}_{a}&=&-2\frac{16!}{13!}A^{IJ}_{a},\\
c^{}_2(a^{}_{4,7}A^{IJ}_{a}+b^{}_{4,7}B^{IJ}_{a})+c^{}_4b^{}_{8,7}C^{IJ}_{a}&=&-2\frac{16!}{9!}A^{IJ}_{a}.
\eea
After numerical evaluation of the coefficients $a_{n,k},b_{n,k},c_{n,k}$, and $d_{n,k}$,  we find,

\be
C^{IJ}_{a}=-\frac{3}{4}(416A^{IJ}_{a}-B^{IJ}_{a}),\qquad D^{IJ}_{a}=-2520(3744A^{IJ}_{a}+B^{IJ}_{a}),
\ee
which enables us to express the commutator in terms of $A^{IJ}$ and $B^{IJ}$. The commutator turns out to be proportional to  
 
\bea
[\,\delta_1^{}\,,\,\delta^{}_2\,]\,X^{}_{A\,a}&=&(\ov \epsilon_2^I\gamma^\mu_{}\epsilon_1^I){\mc D}^{}_\mu X^{}_{A\,a}+\nonumber\\
&&+\epsilon^{IJ}_{}\frac{1}{768}\left(k_1-2(k_2-780(k_3-6k_4))\right)\left(1632A^{IJ}_{a}+B^{IJ}_{a}\right),\nonumber\\
\eea
so that the desired $A^{IJ}_a$ term always comes accompanied with the unwanted $B^{IJ}_a$. We conclude that  the commutator can be written as a gauge transformation, but, unlike the BLG case, its gauge parameter is independent of the structure function $f^{abcd}$. 

\section{Conclusions}
We have shown by two different techniques that there is no interacting superconformal theory in three dimensions with $128$ fermions and bosons, and sixteen supersymmetries. We used the light-cone superspace with eight Grassmann variables, and its constrained chiral superfield, which in four dimensions  describes ${\mc N}=8$ Supergravity. 

It is still possible that in six dimensions, there exists an interacting superconformal theory with $256$ degrees of freedom. That such a free theory exists had been noted earlier by Hull\cite{HULL}. Our result seems to suggest that it also does not have an interacting analog.

\section{Acknowledgements}
We thank Lars Brink and Sung-Soo Kim for useful discussions at various stages of this work. 
PR  thanks the Aspen Center for Physics for its hospitality, where part of this work was done, and DB thanks Jonathan Bagger for helpful discussions.
This research is partially supported by the Department of Energy Grant No. DE-FG02-97ER41029.


\end{document}